\documentclass[prl,aps,twocolumn,showpacs]{revtex4}
\usepackage{graphicx}
\usepackage{amsmath}
\usepackage{bm}
\usepackage{amstext}
\usepackage{amsxtra}

\usepackage{times}

\begin{document}

\newcommand{\To}{T_c^0}
\newcommand{\kB}{k_{\rm B}}
\newcommand{\dT}{\Delta T_c}
\newcommand{\lo}{\lambda_0}
\newcommand{\cs}{$\clubsuit$}
\newcommand{\thold}{t_{\rm hold}}
\newcommand{\Nmf}{N_c^{\rm MF}}
\newcommand{\Tmf}{T_c^{\rm MF}}
\newcommand{\el}{\gamma_{\rm el}}
\newcommand{\Rcl}{R_0^{\rm (cl)}}
\newcommand{\RB}{R_0^{\rm (B)}}
\newcommand{\tT}{\tau_{T}}

\title{Condensation dynamics in a quantum-quenched Bose gas}

\author{Robert P. Smith, Scott Beattie, Stuart Moulder, Robert L. D. Campbell, and Zoran Hadzibabic}
\affiliation{Cavendish Laboratory, University of Cambridge, J.~J.~Thomson Ave., Cambridge CB3~0HE, United Kingdom}

\begin{abstract}
By quenching the strength of interactions in a partially condensed Bose gas we create a ``super-saturated" vapor which has more thermal atoms than it can contain in equilibrium. Subsequently, the number of condensed atoms ($N_0$) grows even though the temperature ($T$) rises and the total atom number decays.
We show that the non-equilibrium evolution of the system is isoenergetic and for small initial $N_0$ observe a clear separation between $T$ and $N_0$ dynamics, thus explicitly demonstrating the theoretically expected ``two-step" picture of condensate growth. For increasing initial $N_0$ values we observe a crossover to classical relaxation dynamics.
The size of the observed quench-induced effects can be explained using a simple equation of state for an interacting harmonically-trapped atomic gas.

\end{abstract}

\date{\today}



\maketitle

Non-equilibrium dynamics of interacting quantum systems are generally far less understood than the corresponding equilibrium many-body states \cite{Polkovnikov:2011}.
Of particular interest are many-body dynamics of both the order parameter and the excitations in a system close to a phase transition.
From a theoretical point of view, a clean and well defined way to induce and study non-equilibrium quantum dynamics is a rapid ``quantum quench" \cite{Calabrese:2006} of a single Hamiltonian parameter. Ultracold atomic gases are very well suited for such quantum quench experiments. In addition to the possibility to dynamically vary microscopic Hamiltonian parameters, they feature near-perfect isolation from the environment and characteristic many-body timescales (ranging from milliseconds to seconds) that are experimentally resolvable and allow real-time non-equilibrium studies.

In this Letter, we introduce a quantum quench of the interaction strength in an atomic Bose gas
as a tool to study the dynamics of Bose-Einstein condensation \cite{Snoke:1989,Stoof:1991,kagan:1992, Semikoz:1995,Griffin:1995,Gardiner:1997a,Gardiner:1998a, Miesner:1998b,Kohl:2002,Ritter:2007,Hugbart:2007,Weiler:2008, Garrett:2011}.
Earlier experiments highlighted the importance of bosonic stimulation in condensate formation \cite{Miesner:1998b}, but could not quantitatively address the theoretically debated interplay of energy redistribution and coherence development in the system \cite{Snoke:1989,Stoof:1991,kagan:1992,Semikoz:1995,Griffin:1995,Gardiner:1997a,Gardiner:1998a}.
The use of a quantum quench of the interaction strength allows us to study these two processes in parallel.
The quench induces a growth of the condensed atom number in a degenerate gas without any removal of thermal energy; we
explain this effect with a simple theoretical model and experimentally study its real-time dynamics.
We explicitly show that the post-quench non-equilibrium evolution of the system is isoenergetic, and directly reveal the theoretically postulated ``two-step" picture of condensation \cite{Stoof:1991,kagan:1992,Semikoz:1995,Griffin:1995}. As expected, close to the critical point the growth of the condensed atom number lags behind the energy redistribution in the thermal component of the gas.
Moving away from the critical point, we also observe a crossover to effectively one-step condensation dynamics governed by a classical relaxation process.

\begin{figure} [bp]
\includegraphics[width=\columnwidth]{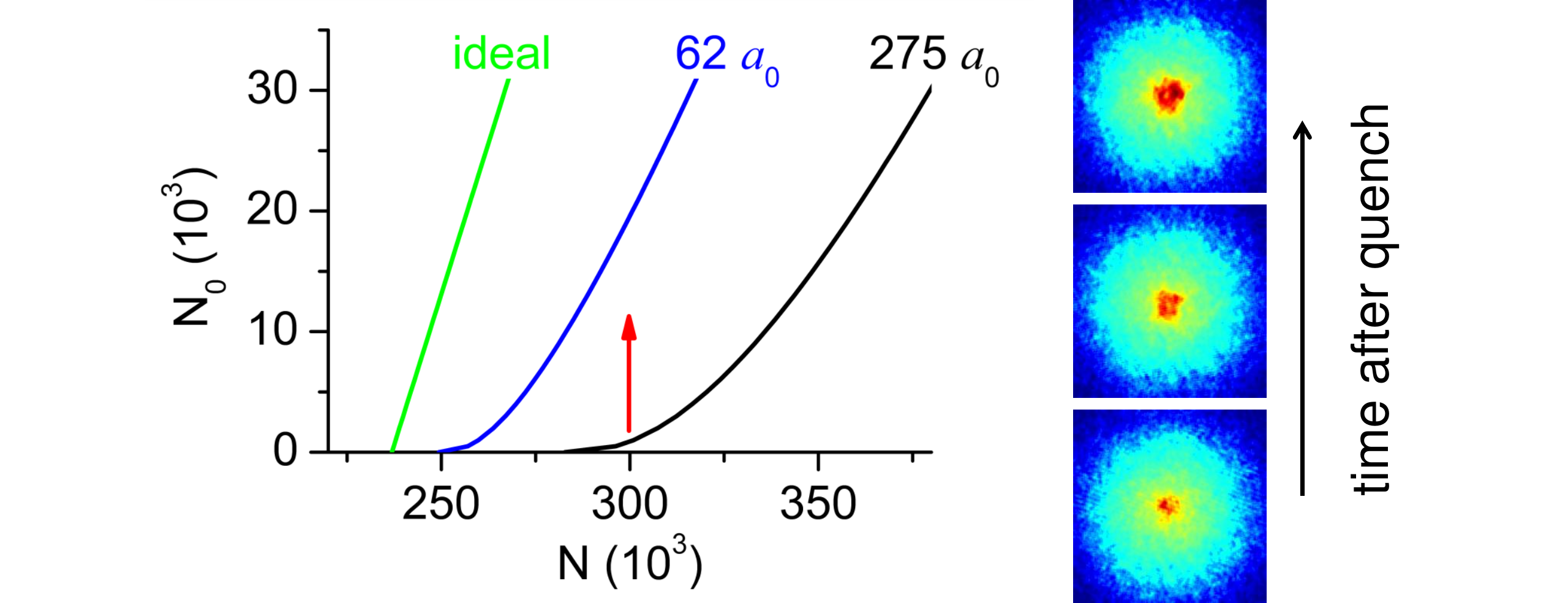}
\caption{(Color online) Inducing non-equilibrium dynamics by a quantum quench of the interaction strength. The equilibrium condensed atom number $N_0$, calculated according to Eq.~(\ref{eq:N}), is plotted versus the total atom number $N$ for a fixed temperature $T =200\,$nK, our trapping parameters, and different scattering lengths $a$. The red arrow indicates the direction of the quench.
The three absorption images of atomic clouds released from the trap show the growth of the condensate (appearing in red) over $\approx 1\;$s following an $a = 275 \rightarrow 62\,a_0$ quench.
}
\label{fig:EoS}
\end{figure}

In an ideal Bose gas the number of condensed atoms, $N_0$, depends only on the total atom number $N$ and the temperature $T$.
In a partially condensed cloud at a given $T$ the number of atoms in the thermal component, $N'$, is saturated at the critical value for condensation, $N_c(T)$, and we have $N_0 = N - N_c$. In experiments on harmonically trapped atomic gases the interactions, characterized by the s-wave scattering length $a$,
change this picture in two ways (see Fig.~\ref{fig:EoS}). First, they induce a shift of the critical value $N_c (a, T)$, which was accurately measured in \cite{Smith:2011}. Second, the thermal component is not saturated - the presence of the condensate allows $N'$ to grow above $N_c$ \cite{Tammuz:2011}. Taking these effects into account, near the critical point we can write the equation of state for an interacting atomic gas in thermal equilibrium \cite{Smith:2012}:
\begin{equation}
N = N_c + S_0 N_0^{2/5} + N_0 \; .
\label{eq:N}
\end{equation}
Here $S_0 N_0^{2/5}$ is the additional number of atoms accommodated in the thermal component due to non-saturation effects, i.e.,  $N'=N_c + S_0 N_0^{2/5}$. The non-saturation coefficient $S_0 \propto a^{2/5} T^2$ can be calculated using mean-field theory \cite{Tammuz:2011, Smith:2012}.
(In our experiments $S_0 \sim 10^3$.)

As illustrated in Fig.~\ref{fig:EoS}, the fact that the equilibrium number of condensed atoms (at a fixed $N$ and $T$) depends on the strength of interactions opens the possibility to use a quantum quench of the scattering length to induce non-equilibrium $N_0$ dynamics.
Here we plot the solutions of Eq.~(\ref{eq:N}) for a fixed $T$ and different scattering lengths $a$, where $a_0$ is the Bohr radius. If the system is prepared in equilibrium at a high scattering length and then $a$ is quenched to a lower value the gas becomes super-saturated, having too large $N'$ and too small $N_0$.
Consequently $N_0$ must grow even without any active cooling of the gas.
Compared to the classical quench of the thermal energy \cite{Miesner:1998b,Kohl:2002,Ritter:2007,Hugbart:2007,Weiler:2008}, the quantum quench of $a$ has the advantage that it does not directly affect $N$ and the {\it initial} thermal occupations of the single-particle excited states. This allows us to study in parallel the induced non-equilibrium evolution of both the coherence and the quasi-thermal energy distribution in the system.

For our experiments we use an optically trapped cloud of $^{39}$K atoms in the $|F,m_F\rangle = |1,1\rangle$ hyperfine ground state \cite{Campbell:2010}, in which $a$ can be tuned via a Feshbach resonance centred at 402.5 G \cite{Zaccanti:2009}. The geometric mean of the harmonic trapping frequencies in our nearly isotropic trap is $\bar{\omega}/2\pi \approx 75\,$Hz, the temperature of our clouds is $T \approx 200\,$nK, and the total atom number $N \approx (3-4) \times 10^5$.

\begin{figure} [tbp]
\includegraphics[width=1.0\columnwidth]{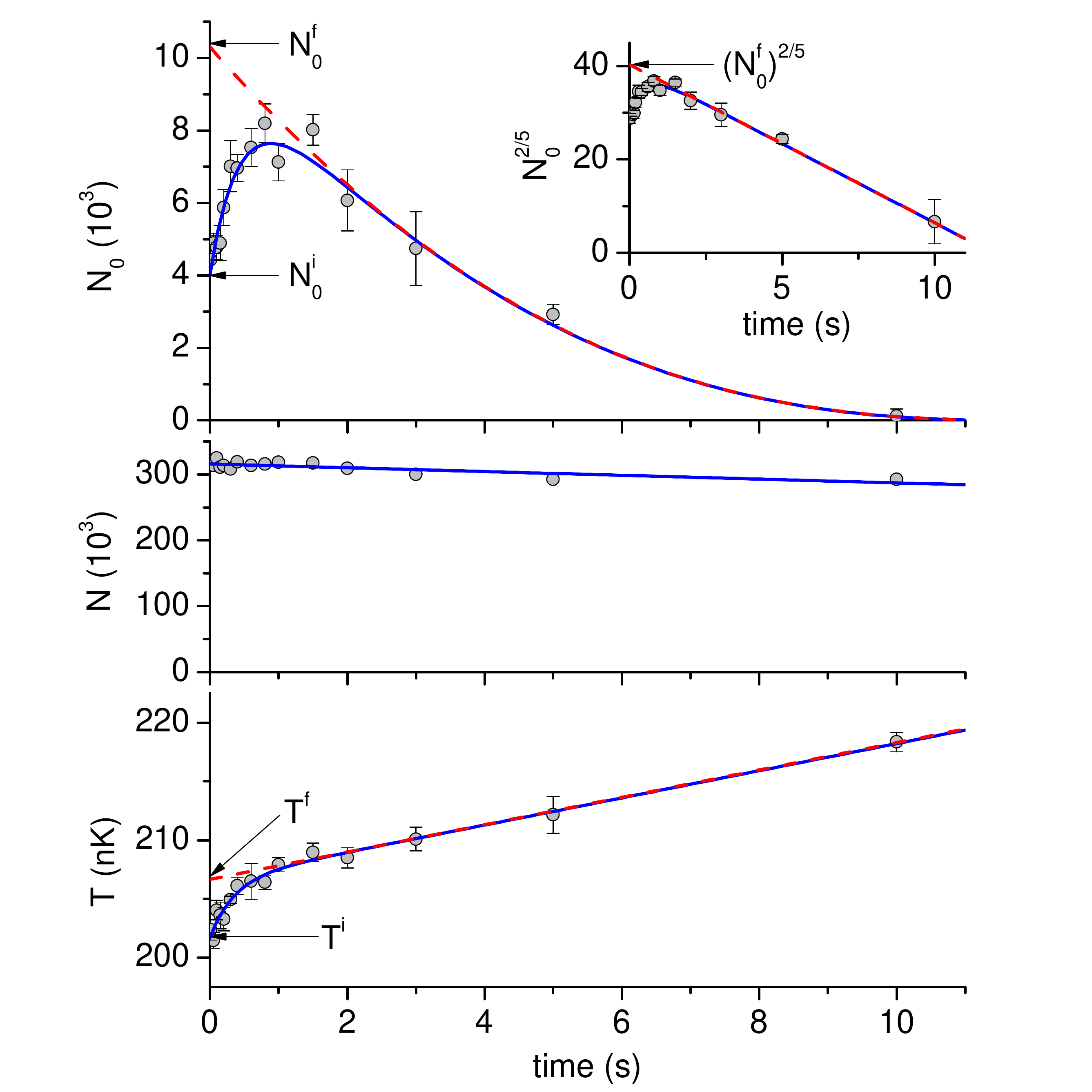}
\caption{(Color online) Non-equilibrium dynamics following an interaction quench from $a^i = 275\,a_0$ to $a^f = 62\,a_0$.  The condensed atom number $N_0$, total atom number $N$, and temperature $T$ are plotted versus the time $t$ after the quench. Each data point is an average of 3 - 7 experimental shots and the error bars are statistical. $N_0$ first grows towards the new equilibrium and then slowly decays due to the slow $N$ decay and a small background heating rate of $\approx 1\;$nK/s. The $N$ decay has a characteristic timescale of $\sim 100\,$s and is essentially linear over 10~s.
In addition to the constant background heating rate, the temperature shows a fast initial rise induced by the quench. The inset shows the evolution of $N_0^{2/5}$, which decays linearly at long $t$. The dashed red lines show fits to the long-$t$ data, used to extract $N_0^f$ and $T^f$ (see text).
}
\label{fig:Quench}
\end{figure}

We initially prepare an equilibrium cloud just below the condensation temperature at a scattering length $a^i = 275\,a_0$ and then rapidly quench the scattering length to a lower final value, $a^f$, by changing the externally applied magnetic field over $\tau_{\rm q} = 10\,$ms. After the quench we follow the evolution of the system for times up to $t=10\,$s, extracting $N_0$, $N$ and $T$ from absorption images taken after 18~ms of time-of-flight (TOF) expansion from the trap \cite{T}. An example of such a data series, with $a^f = 62\,a_0$ and an initial condensate atom number $N_0^i \approx 4 \times 10^3$, is shown in Fig.~\ref{fig:Quench}. We collected a total of 20 such experimental series, each with a different combination of $a^f$ in the range $52 - 97\,a_0$ and $N_0^i$ in the range $(1-40) \times 10^3$.

The quench time $\tau_{\rm q}$ and the $a^f$ values simultaneously satisfy several requirements:

(i) Even for fixed $N_0$ the spatial size of the condensate depends on $a$. Our $\tau_{\rm q}$ is sufficiently long compared to the trap time ($\bar{\omega} \tau_{\rm q} \approx 5$) to allow adiabatic adjustment of the condensate shape at the initial $N_0$ \cite{Matthews:1998}.

(ii) At the same time the quench must be diabatic with respect to the exchange of particles between the thermal cloud and the condensate, so we choose $a^f$ values small enough for the elastic collision rate $\el$ \cite{gamma} to be much smaller than $1/\tau_{\rm q}$.

(iii) The $a^f$ values are large enough for the system to converge towards new equilibrium at long times, rather than forever remaining in an intrinsically non-equilibrium state \cite{Smith:2011, Smith:2012}.

As shown in Fig.~\ref{fig:Quench}, the number of condensed atoms $N_0$ clearly grows following the quench, before eventually decaying at much longer times due to the mundane reasons of slow $N$ decay and a background heating rate of about $1\,$nK$/$s \cite{loss}.
Since at short times $N_0$ grows while $N$ decays and $T$ rises, this increase in the number of condensed atoms is unambiguously an interaction effect.
Interestingly, the temperature also shows a fast initial rise which is clearly associated with the interaction quench.

In order to quantitatively study the non-equilibrium effects that occur on short timescales ($\lesssim 1\;$s) after the quench, we need to eliminate from our analysis the long-term (quasi-static)  drifts of $N$ and $T$. Specifically, we need to experimentally extract the ``target" final values $N_0^f$ and $T^f$ that the system would tend to in absence of the slow background heating and atom number decay (see dashed red lines in Fig.~\ref{fig:Quench}).
The target temperature $T^f$ is simply determined by subtracting the constant background heating rate, but the extraction of $N_0^f$ is a bit more subtle. The slow linear decay of $N$ and rise of $T$ both to leading order correspond to a linear decrease of $N - N_c$ with time. Further, for small condensates, the $S_0 N_0^{2/5}$ term in Eq. (\ref{eq:N}) is significantly larger than $N_0$. Hence, at long times $N_0^{2/5}$ decays linearly, allowing us to extract $(N_0^f)^{2/5}$ by linear extrapolation to $t=0$.

We qualitatively anticipated the quench-induced growth of $N_0$ by solving Eq.~(\ref{eq:N}) under the constraint of fixed $T$ (see Fig.~\ref{fig:EoS}). However we can also understand why the quench must lead to an increase in the temperature of the cloud. For our experimental parameters the average energy of thermal atoms is approximately $3 k_BT \approx h \times 13\,$kHz, while the energy of the condensed atoms (including kinetic, potential, and  interaction energy at $a^f$) is less than $h \times 1\,$kHz. Therefore, during the non-equilibrium evolution of the system after the quench, the atoms moving from the thermal cloud into the condensate take with them much less than ``their share" of energy.
Hence an isothermal non-equilibrium evolution would not conserve energy.
To explicitly test whether the evolution is isoenergetic, we make a simple prediction that relates the changes
$\Delta T = T^f - T^i$ and $\Delta N_0 = N_0^f - N_0^i$ under the constraint of constant energy.
In analogy with the standard models of evaporative cooling \cite{Luiten:1996, kett96evap}, we get that to leading order (for small condensed fractions)
the small fractional increase in $T$ should be equal to the increase in the condensed fraction:
\begin{equation}
\frac{\Delta T}{T} \approx \frac{\Delta N_0}{N} \; .
\label{eq:evap}
\end{equation}

\begin{figure} [btp]
\includegraphics[width=\columnwidth]{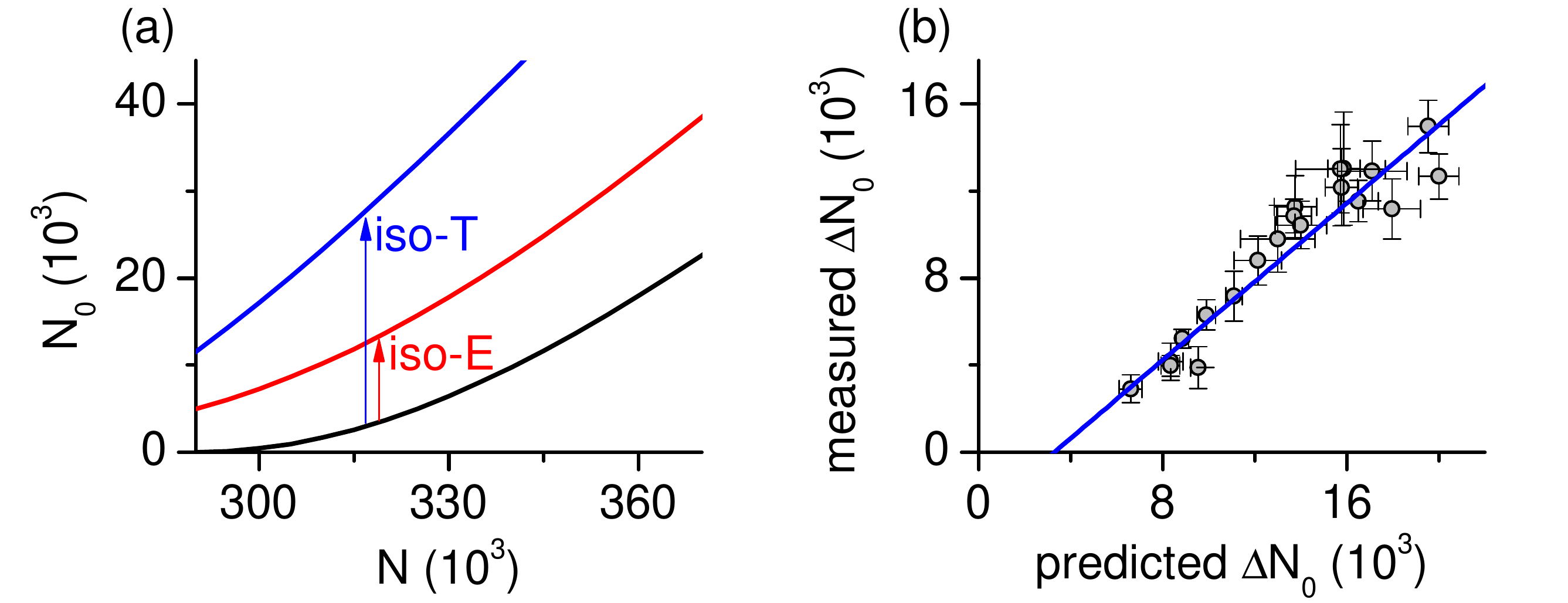}
\caption{(Color online) Isoenergetic non-equilibrium evolution. (a)  For a quench from $T^i=200\,$nK, and $a^i=275 \, a_0$  to $a^f=62 \, a_0$, the three lines show the calculated $N_0^i$ (bottom, black), $N_0^f$ in the isoenergetic picture (middle, red) and $N_0^f$ in the isothermal picture (top, blue). (b) Measured $\Delta N_0$, for 20 experimental series with various $N_0^i$ and $a^f$ values, is plotted versus the predictions of the isoenergetic model. The liner fit to the data (solid blue line) gives a slope of $0.9 \pm 0.1$ and an offset of $(3 \pm 1) \times 10^3$. }
\label{fig:Equilibrium}
\end{figure}

For our 20 data series with different $N_0^i$ and $a^f$ values we observe $\Delta T/T^i = (1.2 \pm 0.3) \Delta N_0/N$, in good  agreement with the prediction of Eq.~(\ref{eq:evap}).
We can now also predict $\Delta N_0$ for any $N_0^i$ and $a^f$, by numerically solving Eq.\,(\ref{eq:N}) under the constraint set by Eq.\,(\ref{eq:evap}) (see Fig.\,\ref{fig:Equilibrium}). As shown in Fig.~\ref{fig:Equilibrium}(b), the measured $\Delta N_0$ follows our theoretical predictions, although we systematically observe slightly smaller $N_0$ increase than predicted \cite{offset}.

\begin{figure} [tbp]
\includegraphics[width=1.0\columnwidth]{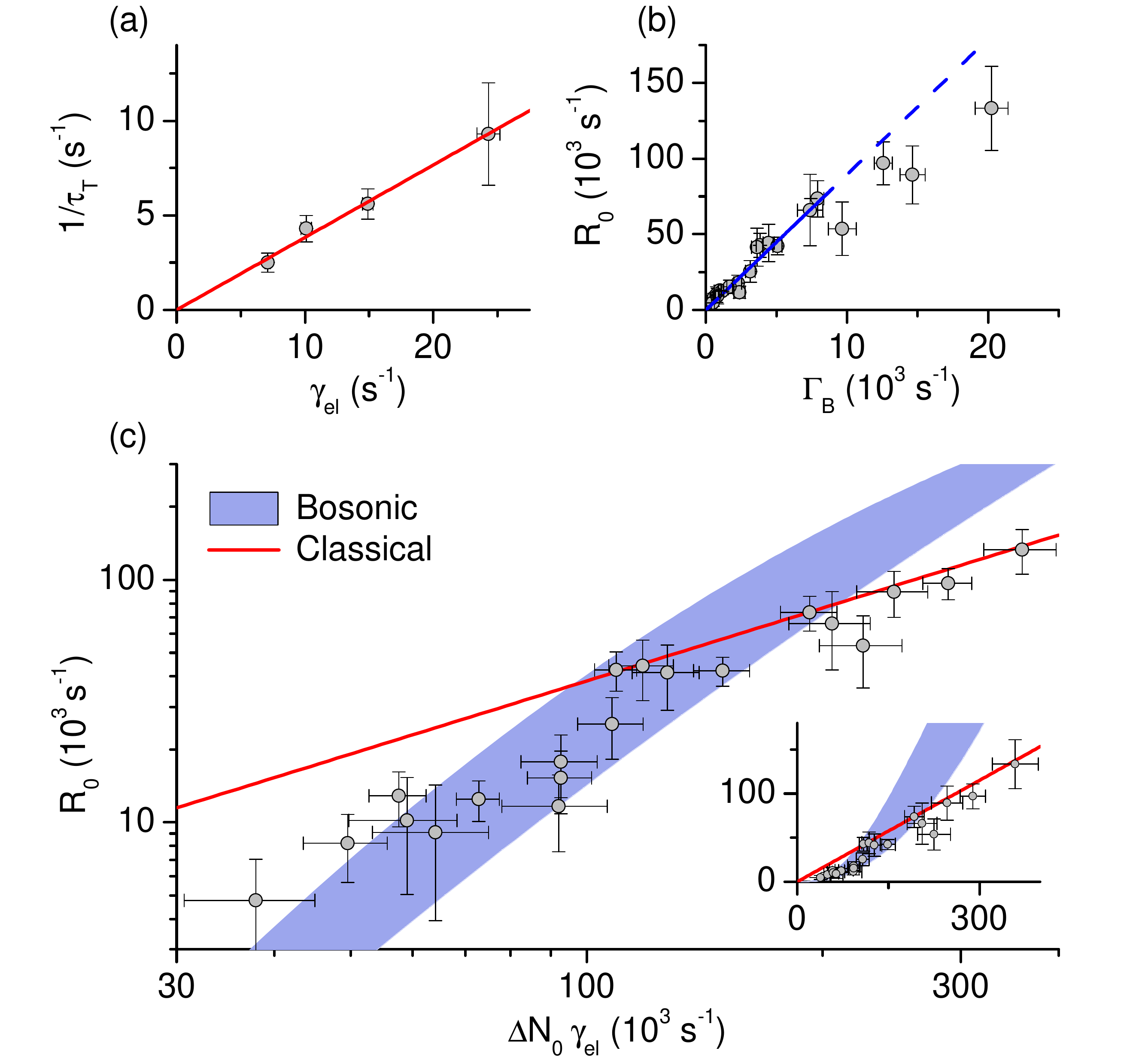}
\caption{(Color online) Condensation dynamics. (a) The $T$ dynamics agree with the classical relaxation picture; the linear fit (red line) gives a relaxation time  $\tT = (2.6 \pm 0.1)/\el$. (b) In the bosonic stimulation picture $R_0 \propto \Gamma_{\rm B}$ (see text).  The linear fit (solid blue line) to low-$R_0$ data gives $\RB = (9 \pm 1)  \Gamma_{\rm B}$. (c)  One- versus two-step condensation. Solid red line shows the one-step classical relaxation prediction, $\Rcl$. The blue shaded area, corresponding to the blue line in (b), shows the two-step bosonic stimulation result, $\RB$.
For increasing $R_0$ we observe a crossover between the two types of dynamics. The inset shows the same data plotted on a linear scale.}
\label{fig:Dynamics}
\end{figure}

In Fig.~\ref{fig:Dynamics} we compare and contrast the evolution of $T$ and $N_0$ during the system's approach to the new equilibrium, for the same 20 data series shown in Fig.~\ref{fig:Equilibrium}(b).

The temperature exhibits classical relaxation dynamics, i.e., $T$ exponentially
approaches $T^f$ on a timescale $\tT$ that depends only on $\el$. As shown in Fig.~\ref{fig:Dynamics}(a),
we find that $\tT$ corresponds to 2.6 collisions per particle (see also, e.g., \cite{Monroe:1993,DeMarco:1999a}). We observe no dependence of $\tT$ on $N_0^i$.

The $N_0$ dynamics are more intriguing.
For $N_0$ to grow, two conceptually distinct steps must take place: (1) the redistribution of the kinetic energy within the gas (seen in the $T$ dynamics) and (2) the merging of the accumulated low-energy atoms into the coherent condensate.
In this two-step picture, (only) the second step depends on $N_0^i$ due to bosonic stimulation, which enhances scattering into an already highly occupied state.
Specifically, the initial $N_0$ growth rate, $R_0 = \dot{N_0}(t=0)$, should be proportional to $\Gamma_{\rm B} = N_0^i \, \el \, \Delta \mu/ (\kB T)$, where $\Delta \mu \sim a_f^{2/5} \,\Delta (N_0^{2/5})$ is the difference between the initial and final chemical potentials \cite{Gardiner:1997a,Gardiner:1998a, Miesner:1998b}.

However, the two-step picture is experimentally relevant only if the second step is the slower, ``rate limiting process", so that $N_0$ growth lags behind the energy redistribution.
If the first step is the rate limiting process, and the particles scattering to low energies essentially immediately join the condensate, then we effectively have a one-step process,
described by the classical relaxation model. In that case we expect $R_0$ to be simply given by $\Rcl = \Delta N_0/ \tT \approx \Delta N_0 \, \el /2.6$.
We qualitatively expect a crossover from the bosonic stimulation (two-step) to the classical relaxation (one-step) behaviour as we move away from the critical point by increasing $N_0^i$. In essence, $R_0$ should always reflect the slower of the two processes.

In Fig. \ref{fig:Dynamics}(b) we plot $R_0$ versus $\Gamma_{\rm B}$ \cite{exponential}.
The low $R_0$ data shows the expected proportionality and the linear fit (solid blue line) gives $\RB = (9 \pm 1)  \Gamma_{\rm B}$.
For large $\Gamma_{\rm B}$ values we see a systematic downwards deviation from this fit. This is what is expected once $\RB$ exceeds the rate at which the classical relaxation of the thermal component feeds atoms into the low-energy states.
In this regime the experimentally observed $R_0$ should be lower than $\RB$ and given by $\Rcl$.

In Fig. \ref{fig:Dynamics}(c) we directly compare the two-step and one-step pictures, and show the crossover between the two types of behaviour. Here we re-plot the $R_0$ data versus $\Delta N_0 \,  \el$, so that the one-step classical result, $\Rcl$, corresponds to the straight line shown in red. The bosonic stimulation result, $\RB$, is now not a universal curve, since it depends on a different set of parameters; the blue line in Fig. \ref{fig:Dynamics}(b) here maps onto the blue shaded area.
We now explicitly see that the small $R_0$ data lies systematically below the classical relaxation prediction, as expected in the two-step picture.
However we also see that the data is consistent with both theories in the crossover region where $\RB \approx  \Rcl$, and eventually agrees better with $\Rcl$ for the largest $R_0$ values.

In conclusion, we have used a quantum quench of the interaction strength to create a super-saturated non-equilibrium Bose gas and study its dynamics.
We have shown that the non-equilibrium evolution of the system is isoenergetic and that the quench-induced changes in the condensed fraction and temperature of the gas can be accounted for using a simple equation of state for an interacting gas in thermal equilibrium.
Moreover, for the first time we directly compare and contrast the energy-distribution and coherence dynamics, and
clearly resolve the two theoretically expected steps in the condensation process.
Here we focused on the case of small but non-zero initial condensates;
with this case understood, in the future it should be possible to use a similar quantum quench to drive the system through the critical point. In that case, in absence of the initial condensate ``seed", it should be possible to study the stochastic effects associated with the spontaneous symmetry breaking and initial condensate formation.


We thank R. Fletcher and A. Gaunt for comments on the manuscript and N. Tammuz for experimental assistance.   This work was supported by EPSRC (Grants No. EP/G026823/1 and No. EP/I010580/1) and a grant from ARO with funding from the DARPA OLE program.


\begin{thebibliography}{10}
\bibitem{T}
In the initial non-equilibrium regime, at short times after the quench, $T$ is strictly speaking not defined. However, in our experiments the quench results in changes of the apparent temperature by $< 5\%$. For such small deviations, at all times an equilibrium thermal distribution fits the data very well and provides a good measure of the thermal energy. For fitting the effective temperature we exclude the central portion of the cloud corresponding to one thermal radius.
\bibitem{gamma}
$\el= \sqrt{2} \langle n  \rangle v_{\rm th} \sigma_0$, where $\langle n  \rangle$ is the average density of thermal atoms at the critical point, $v_{\rm th}=\sqrt{8\kB T/\pi m}$ and $\sigma_0=8\pi a^2$.
We calculate $\langle n  \rangle$ using an ideal Bose gas distribution at the critical point.
\bibitem{loss}
The main sources of the slow atom loss and the background heating are off-resonant spontaneous scattering of photons from the trapping laser beams and small fluctuations in the trapping laser power. The one-body loss due to collisions with the background gas in the vacuum chamber has an even longer characteristic timescale of $\sim 400\;$s.
\bibitem{exponential}
To experimentally extract $R_0$ we heuristically use an exponential fit to the short-$t$ data, taking into account the slow long-term decay of $N_0$. In the bosonic stimulation picture this functional form is not exact, but within experimental precision it fits the data with non-zero $N_0^i$ sufficiently well. For example, the solid blue line in Fig.~\ref{fig:Quench} is based on such a fit.
\end{thebibliography}

\end{document}